\documentclass[runningheads]{llncs}
\usepackage[T1]{fontenc}
\usepackage{booktabs}
\usepackage{url}
\usepackage{graphicx}
\usepackage{xcolor}
\usepackage{hyperref}
\hypersetup{
    colorlinks=true,
    linkcolor=blue,
    urlcolor=blue,
    citecolor=blue
}
\usepackage{color}

\begin{document}
\title{A Survey of Passive Sensing for Workplace Wellbeing and Productivity}
%
%
\author{Subigya K. Nepal\inst{1}\textsuperscript{$\dagger$}\orcidID{0000-0002-4314-9505} \and
Gonzalo J. Martinez\inst{2}\textsuperscript{$\dagger$}\orcidID{0000-0002-7889-4791} \and
Arvind Pillai\inst{3}\orcidID{0000-0002-2489-1130}\and
Koustuv Saha\inst{4}\orcidID{0000-0002-8872-2934}\and
Shayan Mirjafari\inst{3}\orcidID{0000-0002-7165-2781}\and
Vedant Das Swain\inst{5}\orcidID{0000-0001-6871-3523}\and
Xuhai Xu\inst{6}\orcidID{0000-0001-5930-3899}\and
Pino G. Audia\inst{3}\orcidID{0000-0001-7299-513X}\and
Munmun De Choudhury\inst{7}\orcidID{0000-0002-8939-264X}\and
Anind K. Dey\inst{8}\orcidID{0000-0002-3004-0770}\and
Aaron Striegel\inst{2}\orcidID{0000-0002-3157-2859}\and
Andrew T. Campbell\inst{3}\orcidID{0000-0001-7394-7682}}
\authorrunning{S. Nepal and G. Martinez et al.}
%
\institute{Stanford Institute for Human-Centered Artificial Intelligence, Stanford CA 94305, USA\\
\email{sknepal@stanford.edu}\\ \and
University of Notre Dame, Notre Dame IN 46556, USA\\
\email{gmarti11,striegel@nd.edu}
\and
Dartmouth College, Hanover NH 03755, USA\\
\email{arvind.pillai.gr@dartmouth.edu, shayan, campbell@cs.dartmouth.edu}\and
University of Illinois Urbana-Champaign, Urbana IL 61820, USA\\
\email{ksaha2@illinois.edu}\and
Northeastern University, Boston MA 02115, USA\\
\email{v.dasswain@northeastern.edu}\and
Columbia University, New York NY 10027, USA\\
\email{xx2489@columbia.edu}\and
Georgia Institute of Technology, Atlanta GA 30332, USA\\
\email{munmund@gatech.edu}\and
University of Washington, Seattle WA 98195, USA\\
\email{anind@uw.edu}
}

\maketitle  

\vspace{-1em}
\begin{center}
\small\textcolor{red}{\textit{Published at HCI International 2025, DOI: \url{https://doi.org/10.1007/978-3-031-93845-0_29}}}
\end{center}
\vspace{0.5em}
%
\let\thefootnote\relax\footnotetext{$\dagger$ These authors contributed equally to this work.}
\begin{abstract}
The modern workplace is undergoing a radical transformation, driven by technological advances that blur the boundaries between human capability and digital augmentation. At the forefront of this evolution is passive sensing technology - a suite of tools that quietly monitor and interpret human behavior without active user engagement. This paper examines how these technologies are reshaping our understanding of workplace dynamics, with a particular focus on employee wellbeing and productivity. Through a comprehensive review of recent research, we explore both the transformative potential and inherent challenges of passive sensing in professional environments. Our analysis reveals emerging patterns in how these technologies can support worker health and performance, while also highlighting critical gaps in current research and opportunities for future innovation. We conclude by outlining a roadmap for integrating passive sensing into future workplaces in ways that enhance human potential while preserving dignity and autonomy.

\end{abstract}

%
%
%
\section{Introduction}
As technology becomes increasingly integrated into the workplace, researchers studying the ``Future of Work" are exploring new ways to enhance work environments and support employee capabilities. Among emerging technologies, passive sensors have shown particular promise for their ability to blend seamlessly into work settings. These sensors gather data without requiring active input from users, offering an unobtrusive way to understand and support workers throughout their day \cite{review2018}.

What makes these sensors particularly valuable is their ability to capture detailed insights into worker behavior while enabling evidence-based analysis grounded in real workplace contexts. Building on this capability, researchers have uncovered strong connections between passive sensing data and key behavioral indicators, including mental wellbeing \cite{wang2014studentlife}, personality traits \cite{wang2018sensing}, and productivity \cite{10.1145/3328908}. Traditional performance evaluations often rely on self-reporting or supervisor assessments, which provide only periodic snapshots of worker experience. In contrast, passive sensing technologies can track and understand workers' experiences continuously over long periods and across different settings, offering a more complete picture of workplace dynamics without disrupting normal work patterns.
Given the growing importance of these technologies, we present a comprehensive survey of current research on how passive sensing is being used in workplaces to assess and improve worker wellbeing and productivity. For this study, we define "workplace" as any setting where people perform their job duties, whether in private or shared spaces. This inclusive definition reflects the evolving nature of work environments and ensures our analysis captures the full spectrum of workplace applications.

The significance of this research has been amplified by recent advances in Artificial Intelligence (AI) and Large Language Models (LLM), which have created new opportunities for analyzing and acting on passive sensing data. As organizations adapt to changing workforce needs, combining passive sensing with AI could enable more sophisticated ways to support workers. However, this technological convergence also raises important questions about privacy, ethics, and finding the right balance between technological assistance and human autonomy in tomorrow's workplace. Through our analysis, we identify these key challenges and opportunities, providing a foundation for understanding how passive sensing technologies can thoughtfully be integrated into the future of work.
\begin{figure}[] 
\centering
\includegraphics[width=\linewidth]{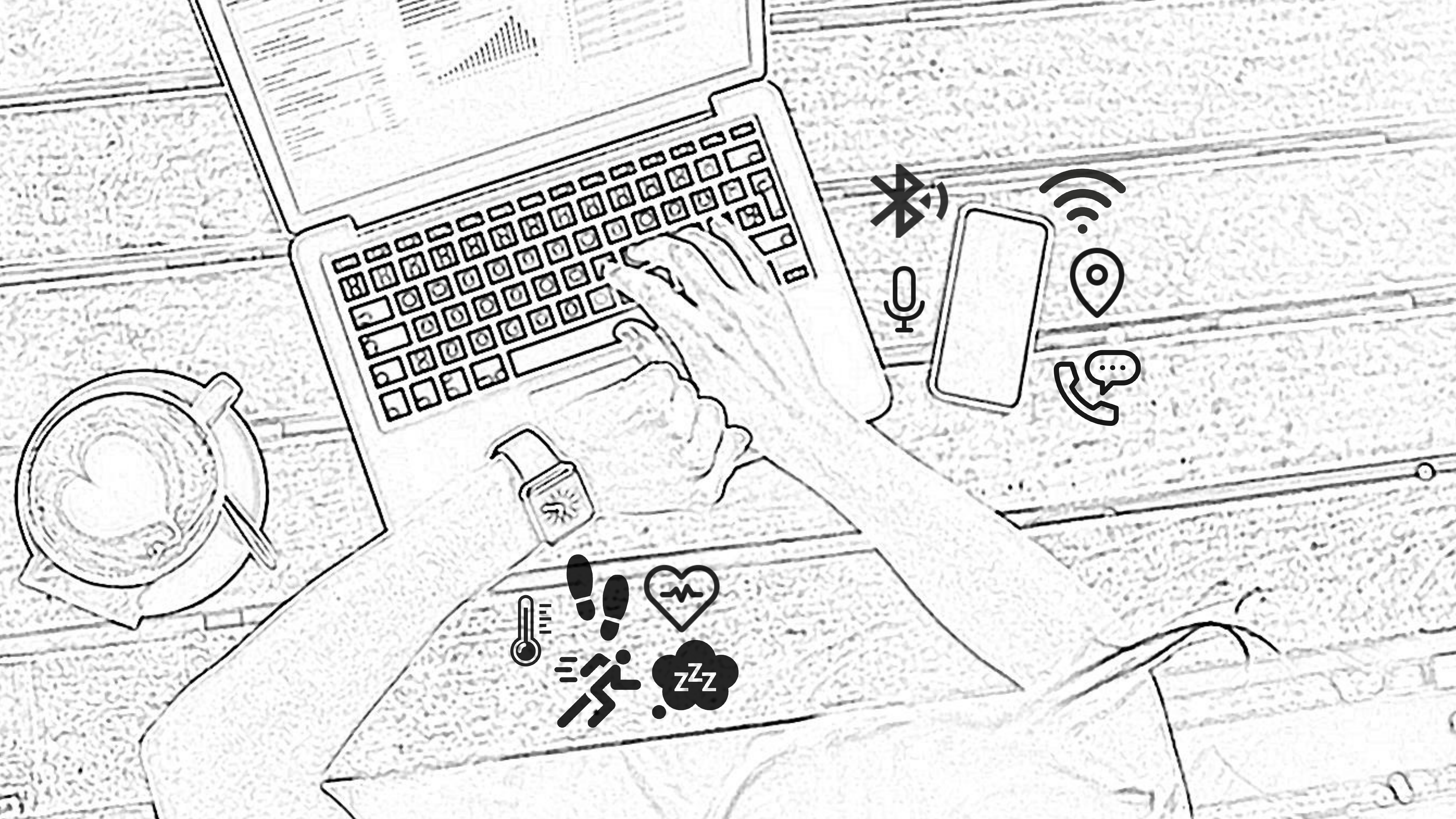}
\caption{Wearables and smartphones are two common passive sensing technologies, offering physiological data such as heart rate, skin temperature, physical activity and sleep and contextual data such as location, proximity and phone usage.}
\label{fig:scatch}
\end{figure}
\section{Survey Methodology}
To ensure the relevance and rigor of our research, we conducted searches for terms like ``passive sensing \& future of work", ``wearables \& worker productivity", and ``wearables \& employee health", leading us to review over 25,000 articles on platforms like Google Scholar. Our selection criteria favored recent papers published since 2015, including some seminal earlier works, and prioritized publications with a Google h5 index of 20 or higher. Furthermore, we imposed specific conditions to refine our focus. Studies had to involve the actual workforce engaged in day-to-day job activities, excluding simulations with non-working groups, such as students in office environments. The technologies considered had to be genuinely passive—unobtrusive, portable, and comfortably integrated into daily routines without requiring active interaction. 
The technologies should blend with the workers' routine, rather than the workers changing their everyday routine to satisfy the study requirements. Therefore, research utilizing intrusive devices like EEG headsets, chest straps, gaze trackers, cameras and other cumbersome scientific devices was excluded in favor of those using more discreet and workplace-appropriate technologies. We focus on devices that we expect to have a higher acceptance in the office environment. This stringent selection process allowed us to accurately portray the state-of-the-art developments, progress, and hurdles in this field, providing a foundation to stimulate and guide further research on passive sensing technologies in the workplace. 

\section{Foundations of Workplace Passive Sensing}
Ubiquitous devices, increasingly embedded in everyday life, now capture daily behavior at high resolution through multiple sensors. Researchers have leveraged passive sensing techniques to model users’ behavior and contexts. Even before smartphones and wearables, cell phone sensors (e.g., GPS, Bluetooth) were used for passive context sensing. RealityMining~\cite{eagle2006reality} employed location and Bluetooth data to identify social patterns, socially significant places, and organizational rhythms. CenceMe~\cite{miluzzo2007cenceme} expanded this approach by integrating cameras, microphones, accelerometers, GPS, and other sensors to detect user activities and habits, then shared these on social networks. Beyond personal behavior, passive sensing also enabled large-scale people-centric urban sensing, as MetroSense~\cite{eisenman2006metrosense} combined mobile and static sensors throughout a city to support various applications. As mobile and wearable devices advanced, longitudinal tracking became feasible, demonstrated by studies like StudentLife~\cite{wang2014studentlife} which monitored college students’ behaviors for over two months, assessing mental health, academic performance, and behavioral trends. Passive sensing has since extended into workplaces to monitor physiological factors, improve safety, and enhance efficiency~\cite{khakurel2018tapping}. For example, Tesserae~\cite{mattingly2019tesserae} studied over 700 information workers to measure workplace performance, psychological traits, and physical characteristics. In the rest of the article, we focus on applications of passive sensing technologies in the workplace to assess wellbeing and productivity of the workforce as the transition to the Future of Work takes place.  

\begin{table*}[]
\centering
\caption{Studied behavior and sensors used. The following table lists all the behaviors studied in the papers we surveyed and the corresponding sensor data used.}
\label{tabl:sensors}
\resizebox{\textwidth}{!}{%
\begin{tabular}{@{}llll@{}}
\toprule
\textbf{Studied  behavior} & \multicolumn{3}{l}{\textbf{Sensor/data}}                                                                                                   \\ \midrule
Stress/anxiety             & \multicolumn{3}{l}{\begin{tabular}[c]{@{}l@{}}HR, step count \cite{feng2020modeling}; HR, electrodermal activity, skin temperature \cite{jebelli2019application}; HR, email, calendar, time~\cite{10.1145/3491102.3502027};\\
Skin conductance, skin temperature, acceleration~\cite{Umematsu2020}; workplace behavioral logs/ computer activity~\cite{morshed2022advancing,10.1145/3544549.3585626,nepal2024burnout}, context~\cite{suh2024toward};\\HR, stress, physical activity, context, user state, phone usage, Bluetooth~\cite{robles2020jointly}; HRV, step count, sleep~\cite{10.1145/3514259};\\ HR, HRV, step count, sleep, acceleration, fat burn, cadence, breathing depth, sitting time, Bluetooth~\cite{booth2019toward}\end{tabular}}        
\\
\midrule
Sleep                      & \multicolumn{3}{l}{\begin{tabular}[c]{@{}l@{}}HR, sleep duration, sleep efficiency \cite{feng2019discovering}; phone usage, movement, HR \cite{9156211}; \\ HR, stress, physical activity, context, user state, phone usage, Bluetooth~\cite{robles2020jointly}\end{tabular}} \\ \midrule
Affect/mood &
  \multicolumn{3}{l}{\begin{tabular}[c]{@{}l@{}}Skin conductance, skin temperature, acceleration~\cite{Umematsu2020}; HR, step count, sleep~\cite{10.1145/3514259};\\ Sleep, activity duration \cite{binmorshed2019prediction,mark2016workplace}; computer activity \cite{mark2016workplace},  speech activity, Bluetooth \cite{nadarajan2019speaker}; \\ HR, stress, physical activity, context, user state, phone usage, Bluetooth~\cite{robles2020jointly,10.1145/3491102.3502037}; \\ physical activity, HR, skin response, skin  temperature, respiration \cite{Soto2021ObservingAP}; \\ HR, pulse, PPG, ECG, accelerometer, skin temperature \cite{zenonos2016healthyoffice}; Sleep~\cite{10.1145/3555217}\end{tabular}}\\ \midrule
Focus/awakeness            & \multicolumn{3}{l}{Physical activity, HR, skin response, skin  temperature, respiration \cite{Soto2021ObservingAP}; Computer activity~\cite{10.1145/3290605.3300697}} \\ \midrule
Productivity          & \multicolumn{3}{l}{Computer activity \cite{mark2016email,mark2016workplace,8925488,nepal2024user}; Speech, calendar~\cite{8925488}; Sleep~\cite{10.1145/3555217}} \\ \midrule
Task performance &
  \multicolumn{3}{l}{\begin{tabular}[c]{@{}l@{}}HR, step count, Bluetooth \cite{fengmosaic}; Bluetooth \cite{10.1145/3359267}; HR, step count, sleep~\cite{10.1145/3514259}\\ Physical activity, speech activity, face to face interaction, proximity \cite{olguin2009capturing};\\ Distance, location, still duration, sleep duration, sleep dept, phone usage, desk sessions \cite{dasswain2019multisensor}; \\ HR, stress, physical activity, sleep, phone usage, Bluetooth~\cite{10.1145/3328908}; \\ 
  Commute based physiology, weather, commute duration, commute variability~\cite{nepal2021}\end{tabular}} \\ \midrule
Citizenship behavior &
  \multicolumn{3}{l}{\begin{tabular}[c]{@{}l@{}}HR, stress, physical activity, sleep, phone usage, Bluetooth~\cite{10.1145/3328908}; HR, step count~\cite{feng2020modeling,10.1145/3514259};\\
  Distance, location, still duration, sleep duration, sleep dept, phone usage, desk sessions \cite{dasswain2019multisensor}\\
  Commute based physiology, weather, commute duration, commute variability~\cite{nepal2021}\end{tabular}} \\ \midrule
Deviance &
  \multicolumn{3}{l}{\begin{tabular}[c]{@{}l@{}}HR, step count, Bluetooth \cite{fengmosaic};  HR, step count~\cite{feng2020modeling,10.1145/3514259}; \\ HR, stress, physical activity, sleep, phone usage, Bluetooth~\cite{10.1145/3328908}; \\ Distance, location, still duration, sleep duration, sleep dept, phone usage, desk sessions \cite{dasswain2019multisensor} \end{tabular}} \\ \midrule
Promotion   & \multicolumn{3}{l}{Physical activity, HR, stress, phone usage, Bluetooth, sleep, distance~\cite{10.1145/3414118}} \\ \midrule
Cognitive load/interruptibility        &
\multicolumn{3}{l}{\begin{tabular}[c]{@{}l@{}}HR, HRV, EDA, skin temperature, wrist movements~\cite{schaule2018employing}; Speech, computer activity, calendar~\cite{8925488};\\ HR, HRV, step count, sleep, circadian rhythm~\cite{zuger2018sensing}; Computer activity~\cite{mark2016workplace} \end{tabular}} \\ \midrule
Misc. workplace activity & \multicolumn{3}{l}{\begin{tabular}[c]{@{}l@{}}EDA, blood volume pulse (BVP), acceleration, phone usage~\cite{di_lascio}; Computer activity~\cite{10.1145/3290605.3300697,8925488,sefidgar2024improving}; Speech, calendar~\cite{8925488};\\ User activity, application usage, location, ringer mode~\cite{kucukozer2021designing}; Bluetooth~\cite{Clark2018}; Computer-assisted protected time~\cite{das2023focused,saha2023focus} \end{tabular}} \\ \bottomrule
\end{tabular}%
}
\end{table*}

\section{Applications of Passive Sensing in the Workplace}
In this section, we explore how passive sensing technologies advance our understanding of workplace wellbeing and productivity. Drawing on recent research, we highlight their ability to illuminate the intricate relationship between work environments and employee behavior. This exploration underscores the potential of passive sensing to optimize workplace dynamics and promote a healthier, more productive work culture.
\subsection{Understanding Employee Wellbeing}
Over the years, there has been an increase in studies that explore the sensing capabilities of smartphones and wearables to assess as well as improve worker's health and wellbeing. Among the papers selected, mental health issues such as stress, anxiety, affects are important dimensions studied. Most of these studies ask participants to self-report their health and wellbeing using validated instruments. While some studies use one-off survey scores, others employ experience sampling to study worker's wellbeing in a longitudinal fashion. We list the key studies in Table \ref{tabl:sensors}. 

Assessing workers' stress and/or anxiety is perhaps the most studied topic about worker's wellbeing. As stress and anxiety are known to have an impact on heart rate (HR), most of the studies rely on HR-based signals obtained from wrist-worn wearables that employ photoplethysmography (PPG)~\cite{ppg} sensors. For instance, Feng \textit{et al.}~\cite{feng2020modeling} show that PPG based HR and step counts from a Fitbit can be used to classify the anxiety level of nursing professionals. The authors introduce a pipeline for discovering behavioral consistency features, the inclusion of which leads to an overall improvement in the predictive model, with an accuracy of about 58\%. For this work, the authors use the median-splitted score of self-reported State-Trait Anxiety Inventory (STAI)~\cite{stai} as the ground truth. Similarly, in a study of 10 construction workers, Jebelli \textit{et al.}~\cite{jebelli2019application} use a wrist wearable capable of not only capturing the participant's HR through PPG signals, but also their electrodermal activity (EDA) and peripheral skin temperature (ST). The study uses salivary cortisol as the ground truth to assess the baseline stress levels of the workers, obtaining an accuracy of about 73\% when distinguishing between low, medium and high stress level of the workers.  Similarly, Mark \textit{et al.}~\cite{mark2016email} discover in their research on information workers that there is a direct correlation between the daily time spent on emails (captured via activity logging software) and increased levels of reported stress. In a bid to craft a stress-reduction intervention for the workplace, Howe \textit{et al.}~\cite{10.1145/3491102.3502027} conduct a four-week study with 86 participants to evaluate workplace stress-reduction interventions, using signals from email, calendar, facial images, and heart rate data. Their findings show that digital micro-interventions effectively reduce short-term workplace stress. Building on this work, Suh \textit{et al.}\cite{suh2024toward} analyze factors influencing intervention engagement, finding that contextual signals and individual traits significantly impact intervention effectiveness, while user-rescheduled interventions show higher engagement rates. Together, these studies demonstrate the value of passive sensing-based interventions while highlighting the importance of individual and contextual factors in their design.

Sleep quality is another frequently studied health issue. Feng \textit{et al.}\cite{feng2019discovering} study 138 nursing professionals using Fitbit data (PPG-based HR, sleep duration, efficiency, and REM sleep) to predict sleep quality based on PSQI survey\cite{psqi}. Their motif-based approach achieves 77.47\% F1 score, improving upon the 61.90\% baseline using Fitbit summary data alone. Martinez \textit{et al.}~\cite{9156211} examine sleep detection accuracy among 700 information workers, finding that combining phone usage data with wearable data produces better sleep models than using either data source independently. Research has also explored the role of passive sensing in assessing workers' affect, emotions, and mood. Bin Morshed \textit{et al.}\cite{binmorshed2019prediction} analyze mood instability in 603 information workers, finding a negative correlation between sleep/activity duration (from Garmin wearables) and mood instability, measured via self-reported PANAS scores\cite{panas}. Umematsu \textit{et al.}\cite{Umematsu2020} use wrist wearables to track skin conductance, temperature, and acceleration in 39 Japanese workers over 30 days. Their predictive model forecasts next-day stress, mood, and health scores with mean absolute errors of 13.47, 14.09, and 18.51, respectively. For fine-grained mood classification, Zenonos \textit{et al.}\cite{zenonos2016healthyoffice} develop a framework using physiological data (HR, pulse, PPG, ECG, skin temperature) to classify eight moods into five categories. Mark \textit{et al.}~\cite{mark2016workplace} examine workplace email usage and find a negative correlation with mood balance—higher email engagement corresponds to lower positive affect.

In another study of 50 hospital workers which also uses PANAS, Nadarajan \textit{et al.}~\cite{nadarajan2019speaker} find that speech activity can explain some variance in predicting positive affect measure. Employees wear a specifically designed audio badge during their work-shift hours. The authors extract several features from the audio to identify foreground speech. They then use a linear mixed effects model to estimate positive and negative affect from foreground activation (i.e., the percentage of recording time that foreground speech is present). Similarly, in another study, Robles-Granda \textit{et al.}~\cite{robles2020jointly} utilize multiple sensing modalities to assess anxiety, sleep and affect of 757 information workers. Sensing modalities include wearable, phone application, Bluetooth beacons and social media. Models trained on the fusion of all the features from different sensing modalities lead to up to 13.9\% improvement in the symmetric mean absolute percentage error (SMAPE) when predicting affect, anxiety and sleep quality scores. Booth \textit{et al.} employ multiple sensing modalities to study stress, anxiety and affect of hospital workers~\cite{booth2019toward}. Using Bluetooth tags for location tracking, Fitbit data, and smart shirt sensors, authors achieve 58\% accuracy in classifying participants' mental wellness between high and low stress/affect states. Similarly, Nepal \textit{et al.}~\cite{nepal2024burnout} study burnout in cybersecurity incident responders using both passive sensing and self-reports. Analyzing digital activities of 35 security workers, they find that burned-out employees exhibit distinct behavioral patterns - sending more emails after hours, having more weekend work, and showing poorer work-life boundaries.

Other wellbeing-related studies using passive sensing include focus and awakeness. Soto \textit{et al.}\cite{Soto2021ObservingAP} leverage biometric data from an arm-worn device (physical activity, HR, skin response, skin temperature, respiration) to estimate stress, focus, and awakeness in 14 knowledge workers over eight weeks. Their personalized models outperformed generic ones, improving precision, recall, and accuracy by up to 52\%. There are also several other works that fuse passive sensing data with some other modality to assess wellbeing. For instance, Saha and Grover \textit{et al.}~\cite{koustuvpersoncentric} propose contextualizing on offline behaviors as obtained from passive sensors to make models better adapted to the social media signals. Authors first cluster participants based on the passive sensing data obtained from their wearable, Bluetooth beacons and smartphone application. Thereafter, they use social media derived features to predict different constructs within each cluster, obtaining an improvement on the baseline generic model by up to 5.43\% in predicting self-reported anxiety, affect, and sleep quality of information workers. 

Passive sensing technologies are also being embraced by organizations to promote employees' wellbeing. Organizations make use of gamification, personalized recommendations or even offer incentive programs to encourage employees to be more active in their day-to-day life~\cite{khakurel2018tapping}. Researchers studying the use of wearable technologies in corporate wellness programs report that their usage has a positive impact on employee wellbeing and health~\cite{Giddens2017TheRO}. Passive sensing technologies, thus, can not only be used to monitor and assess health and wellbeing of workers, but they may also be utilized for health interventions and promoting a healthy lifestyle for the workers. 

\subsection{Measuring Workplace Performance}
Driven by early successes in assessing organizational wellbeing, researchers have shifted to using passive sensing for workplace performance. By tracking behaviors, they aim to identify factors that boost productivity and develop interventions to improve it. Yet objectively and broadly quantifying workplace performance remains challenging. For example, while Amazon uses smartphone-based sensing to discourage drivers from checking their phones, frequent phone usage might be necessary in a consulting role. Therefore, researchers in the ubiquitous computing community have often turned to job performance inventories developed by psychologists as ground truth to measure perceived workplace performance across organizations and industries in a generalizable manner. Passive sensing is then used to predict inventory scores or categorize individuals into higher and lower performance tiers. Typically, these inventories measure task performance (e.g., Individual Task Proficiency [ITP], In-Role Behavior [IRB]) or behaviors fostering organizational effectiveness (e.g., Interpersonal and Organizational Deviance [IOD-ID, IOD-OD], Organizational Citizenship Behavior [OCB]).

Task performance encompasses duties or actions formally acknowledged and rewarded by management~\cite{viswesvaran_task_performance}. Several studies have tried to predict ITP and IRB as measures of task performance. For instance, Feng \textit{et al.}\cite{fengmosaic} study 84 full-time nurses over ten weeks, using Fitbit and Bluetooth proximity data to investigate how physiological responses vary across different hospital rooms. Combining a measure of ``mutual dependency'' between room use and physiological responses with aggregated features (e.g., min, mean, max, SD) yields adjusted $R^2$ scores of .11 for ITP and .12 for IRB. Another study\cite{olguin2009capturing} tracks 67 nurses wearing sociometric badges that measures physical activity, speech, face-to-face interaction, and proximity-based social networks. Multiple linear regression models using these features predict 49\% of the variance in perceived workload and 63\% of the variance in perceived productivity.

In studies of information workers, several works stand out. Das Swain \textit{et al.}\cite{10.1145/3359267} track 249 individuals over 62 days using Bluetooth beacons to measure desk presence, work hours, and home time. They develop an organizational fit metric based on the alignment of individual routines with workplace norms, finding that routine fit, combined with Big Five Personality traits\cite{bigfive}, explains up to 28\% of the variance in IRB, with a significant positive association (p $<$ 0.05). In another 110-day study of 603 information workers, Das Swain \textit{et al.}\cite{dasswain2019multisensor} introduce ``organizational personas,'' clustering individuals based on Big Five personality data and multimodal activity facets from a fitness tracker, phone agent (location, phone usage), and Bluetooth beacons (desk sessions). They find that activity facets contribute modest but statistically significant variance to ITP. Similarly, Pillai \textit{et al.}\cite{pillai2023rare} use an autoencoder-based approach to detect rare life events from multivariate mobile sensing data in 126 information workers. Mirjafari \textit{et al.}~\cite{10.1145/3328908} examine 554 information workers using a phone agent, beacons, and a Garmin fitness tracker. They apply K-Means clustering on ITP, IRB, OCB, IOD-ID, and IOD-OD scores to classify higher and lower workplace performers. Their findings show that features from fitness trackers alone achieve an AUROC of .72, phone agent features .65, and a combination of both .83, demonstrating the benefits of multimodal sensing in capturing workplace behaviors. In a follow-up study of 298 information workers, Mirjafari \textit{et al.}\cite{mirjafari2021} use autoencoder-generated features from mobile phones and a Garmin fitness tracker to predict daily job performance changes (improvement, decline, or no change) based on ITP, IRB, OCB, IOD-ID, and IOD-OD scores. Their XGBoost model achieves a 75\% F1 score, significantly surpassing the 33\% baseline. They also apply Gradient Analysis\cite{su_one_2019,liu_facial_2014} to provide user-actionable insights, helping individuals identify behaviors that influence performance. Similarly, Mark \textit{et al.}~\cite{mark2016email} analyze email usage among 40 information workers, finding that increased time spent on emails correlates with lower self-reported productivity.

In the behavioral side of workplace performance inventories, IOD-ID and IOD-OD are measures of ``bad'' conduct in the workplace. Behaviors that indicate IOD-ID can involve cursing a co-worker, playing pranks, or making fun of someone. Behaviors that indicate IOD-OD can be tardiness or absenteeism, leaving work early without permission, putting little effort into work, among others \cite{everton_cwb}. On the other hand, OCB is a measure of ``good'' behavior. OCB is composed of behaviors that reflect altruism, conscientiousness, sportsmanship, courtesy and civic virtue \cite{organ_ocb}. Several studies have focused on these measures. Feng and Narayanan \cite{feng2020modeling} show that behavioral consistency features from wearable data improve accuracy by 6\% compared to summary features alone in a study of 97 hospital workers. Nepal \textit{et al.}~\cite{nepal2021} analyze commute-based sensing data from 275 workers to predict OCB and IOD scores, achieving MAEs of less than 10\%. In a previously mentioned study by Feng \textit{et al.} \cite{fengmosaic}, authors demonstrate that workplace physiological responses predict IOD-ID and IOD-OD (adjusted $R^2$ = 0.065 and 0.092). Multiple studies find associations between passive sensing data and workplace behaviors: Mirjafari \textit{et al.} \cite{10.1145/3328908} with IOD and OCB, Das Swain \textit{et al.} \cite{dasswain2019multisensor} with activity patterns, and beacon-based desk routines \cite{10.1145/3359267} with lower IOD, though accuracies remain moderate (ranging from 56.6\% to 62.4\%). The work by Nepal \textit{et al.} \cite{10.1145/3414118} follows a different line of research. Instead of attempting to predict subjective performance measures from objective data, it intends to find what objective data is related to individual success in organizations, i.e., getting promoted. The authors note that if promotion can be detected and it produces objective physiological signals, researchers could learn from these signals and through the study of job promotion -- which could be inferred -- what behaviors could be behind the high performance that granted the promotion to measure them objectively, foster them and replicate them. 

Nurturing collaboration through open offices has become the norm in the modern era and workers face necessary interruptions throughout their workday to efficiently tackle problems. However, there are studies demonstrating that the nature and timing of interruptions negatively affect workplace productivity \cite{bailey2006need,adamczyk2004if} and increase worker stress \cite{mark2008cost}. Therefore, evaluating interruptibility has significant value to companies in terms of cost and time. To continuously measure a person's interruptibility, Zuger \textit{et al.} \cite{zuger2018sensing} study 13 software engineers over two weeks using wearable sensors in addition to keyboard and mouse interaction data. The ground truth in this study is collected using a pop-up question during work which asked the user to rate their interruptibility on a 1-7 Likert scale, and these ratings are further grouped into three states (12|345|67) for classification. Ultimately, they predicted interruptibility accurately (68.3\%) by training a random forest on several features extracted from sensor modalities like heart rate, activity, and sleep.

Other works target efforts to support the design of future interventions in the workplace. Kimani \textit{et al.}~\cite{8925488} develop a conversational agent that assists information workers with task management, scheduling, prioritization, and minimizing distractions. It leverages contextual data to determine optimal intervention times. In a six-day study with 24 participants, the agent analyzed speech, facial imagery, and computer usage, leading many users to become more aware of their work habits and adjust routines for improved productivity and wellbeing. Nepal \textit{et al.}~\cite{nepal2024user} introduce an LLM-powered productivity agent that utilizes computer-based telemetry to provide personalized assistance. They emphasize the importance of user-centric design, adaptability, and balancing personalization with privacy in AI-driven workplace tools. The integration of passive sensing with LLMs represents a major advancement, allowing for more contextually aware interventions, such as detecting when employees struggle with tasks or optimizing deep work periods. However, this convergence also raises concerns about data privacy, algorithmic bias, and the boundaries between AI assistance and workplace surveillance. 

Kadoya \textit{et al.}\cite{su12041544} use wearables to examine the link between emotional states and productivity, aiming to inform future interventions. Mark \textit{et al.}\cite{mark2016workplace} find that affect balance correlates with concentration and productivity, with sleep playing a key role. In another study, Schaule \textit{et al.}\cite{schaule2018employing} develop a smartwatch-based system to predict cognitive load and determine optimal moments for interruptions, while Di Lascio \textit{et al.}\cite{di_lascio} demonstrate how wearables can passively monitor workplace activity, supporting distraction management and break suggestions. Kucukozer-Cavdar \textit{et al.}\cite{kucukozer2021designing} create a model predicting office workers' availability and break preferences using mobile sensing data. Das Swain \textit{et al.}\cite{das2023focused} assess computer-assisted protected (CAP) time in 89 information workers, showing significant benefits for those new to automated scheduling. Similarly, Saha \textit{et al.}~\cite{saha2023focus} further find that CAP enhances relaxation while reducing negative emotions like anger. These studies deepen our understanding of workplace performance and demonstrate ongoing efforts to develop productivity-enhancing interventions. With the rise of LLMs and AI, integrating passive sensing with advanced analytics presents a promising frontier, paving the way for more adaptive and intelligent workplace interventions.

\section{DISCUSSION}
The works discussed in this article follow a wide variety of approaches to solve the problem of assessing wellbeing and productivity through passive sensing \textit{in-situ}. In what follows, we discuss some considerations to take into account for Future of Work research.

\subsection{Considerations for Ground Truth}
While adapting to the Future of Work, researchers need to make additional considerations beyond simply selecting validated survey instruments. For instance, surveys at the workplace could bias participants from being candid because of employer surveillance. However, remote work might mitigate such concerns. Alternatively, many wellbeing constructs can be measured through physiology (e.g., cortisol), but when the workforce is distributed across diverse work spaces, such measures could include unforeseen artifacts which a consistent workplace would otherwise insulate. Similarly, for productivity measures, while task performance is important, for certain job roles deviance measures or citizenship behavior could provide a more appropriate view of performance. Apart from careful selection of ground truth instruments, researchers should collaborate with industrial and organizational psychologists and personnel management teams to learn which forms of estimations are most relevant to them --- whether that be categorical classifications or continuous evaluations.

\subsection{Considerations for Sensor Deployment}
Ideally, the set of constructs that researchers want to investigate drives the choice of sensors. However, especially while estimating new measures, it can be challenging to ascertain a finite combination of sensors to deploy. Moreover, deployments are expensive and challenging. In longitudinal \textit{in-situ} studies, even if issues like privacy concerns and maintenance of the sensing infrastructure can be mitigated, lack of participant compliance can deteriorate the quality of data~\cite{martinez2020quality}. Also, it may not be practical for all researchers to install a large suite of sensors and triage which signals are actually meaningful. In such cases, researchers can consider forms of low-burden sensing such as logging social media behavior~\cite{saha2019libra,dasswain2020culture,dasswain2020social,de2013understanding,saha2021social}. Therefore, we need to select sensors that sustain minimum burden on the user and still provide data consistently. Furthermore, the deployment of passive sensing technologies, as explored in a case study on just-in-time emotional support agents, underscores the importance of addressing boundary misalignments, data ownership, and power dynamics to facilitate successful implementation~\cite{10388088}. These considerations highlight the nuanced challenges of deploying personalized sensing systems in both work and nonwork contexts, emphasizing the need for careful alignment with user values and wellbeing definitions.

\subsection{Considerations for Machine Learning and AI at Work}
The dynamic nature of workplaces makes passive sensing studies heavily reliant on longitudinal, in-situ data, often leading to missing data challenges. Sensors requiring constant skin contact, such as wearables, are particularly prone to gaps due to movement or non-compliance, and missing data is rarely random, making naïve omission approaches prone to bias~\cite{compliance_paper}. Researchers can mitigate this by leveraging complementary data streams for imputation~\cite{Saha2019imputation,9156211}. Additionally, the high-dimensional, time-varying nature of workplace data demands careful model selection, with approaches such as moment statistics, regularity analysis, and auto-encoders offering varying trade-offs in performance, interpretability, and generalizability~\cite{Xu2019a}.

As AI and LLMs become more integrated into workplace sensing, they offer a powerful means of overcoming these challenges by extracting deeper insights from complex sensor data. LLMs, in particular, can enhance interpretability and actionability by translating raw sensing data into natural language explanations, making insights on wellbeing and productivity more accessible to both employees and managers. This ability to contextualize passive sensing data opens the door to more adaptive and personalized workplace interventions. However, the integration of AI in workplace monitoring also introduces ethical risks, including concerns about privacy, bias, and surveillance overreach. Ensuring fairness, transparency, and strong privacy protections is essential to building trust and preventing misuse. Future research should prioritize human-centered AI models that enhance workplace support while safeguarding worker autonomy and data security.

\subsection{Privacy and Ethical Considerations}
 While we primarily highlighted the opportunities and potentials of these technologies and methodologies, these also come at a cost. For example, despite the best intentions, when instrumented at workplaces, these approaches can bear ethical and privacy concerns~\cite{kawakami2023wellbeing,das2023algorithmic,roemmich2023emotion,corvite2023data,kaur2022didn,adler2022burnout,das2024teacher}. There are lingering questions regarding the best practices of these algorithmic inferences for real-world measures and high-risk decisions because these approaches can be misused in ethically questionable ways. Consequently, these approaches and technologies can compromise privacy, defy expectations, and damage the trust between individuals and technologies~\cite{dasswain2020social}. Park \textit{et al.}\cite{park2021human} identify key themes that negatively affect employee perception of AI use in the workplace, and suggest increased transparency and human-in-the-loop interventions as solutions. Table \ref{tabl:sensors} shows that several approaches rely on sensing that occurs most often outside the workplace, e.g., sleep and physical activity, sensing for which there is no conceivable way to acquire consent from all involved parties such as speech activity, or sensing that could be carried out using the sensors in devices that could be provided by employers or fall into Bring-Your-Own-Device (BYOD) policies such as mobile phones. In fact, consenting to a passive sensing study in the workplace is complex because of the associated power dynamics~\cite{kawakami2023wellbeing,chowdhary2023can,das2023algorithmic}. A study by Chowdhary \textit{et al.}\cite{chowdhary2023can} reveals that the passive sensing framework must support meaningful consent that is freely given, reversible, informed, enthusiastic, and specific.

 In fact, employer surveillance and employee's subjective expectation of privacy share a competing relationship~\cite{ghoshray2013employer}, and only a thin line of difference exists in perceiving the same technology as for surveillance or for supportive intervention~\cite{dasswain2020social}. There is a need to ensure that future work applications benefit individuals without sacrificing their privacy or forcing them to adopt behaviors solely on the need for increasing productivity in the workplace according to arbitrary metrics. 
 In fact, the awareness of being observed or ``monitored'' can lead individuals to deviate from their otherwise typical behaviors---or the phenomenon of observer effect or ``Hawthorne'' effect~\cite{saha2024observer,mccambridge2014systematic}. Recent work by Das Swain \textit{et al.}\cite{das2024sensible} investigates worker attitudes toward passive sensing systems, finding that acceptance varies significantly based on what is being sensed and how the data is used. Their work shows that workers are more accepting of systems that track digital time use or physical activity compared to visual or linguistic data, and prefer insights about mental wellbeing over performance metrics. Importantly, they find that workers resist automatic sharing of these insights with others, highlighting the need for worker agency in data sharing. In a different study, Das Swain \textit{et al.}\cite{das2023algorithmic} conduct in-depth interviews with 28 information workers, revealing deeper concerns about workplace sensing. These include the need for clear boundaries between work and personal life, the importance of human interpretation of sensor data, apprehension about continuous performance quantification, and the desire for granular control over data sharing. Together, these studies emphasize the need for guidelines that lead to responsible applications that balance the trade-offs of risks and benefits associated with these data-driven human-centered assessments~\cite{9597452}. 
 
 In addition, these guidelines should prevent the ingraining of biases in automated decisions and performance-evaluation systems, a danger made obvious by early attempts at developing AI tools for recruiting that resulted in biased results~\cite {dastin_2018}. Considering the lack of diversity in the tech industry in the US~\cite{equalemployment}, there is a considerable risk that the training of models based on existing employee data without special consideration of disproportionate representation could result in AI-based tools encoding biased behaviors that are not truly supporting task performance improvements for the companies or well-being benefits for the employees. These aspects should not be taken lightly and researchers should also incorporate metrics of AI fairness~\cite{aifairness2020} and potential harms~\cite{kawakami2023wellbeing} when conducting future of work studies that predict performance outcomes.
 
 \subsection{Limitations and Future Work}
 The existing literature reveals multiple challenges in using data-driven, human-centered assessments for workplace outcomes. Conflicting definitions of “ground-truth” and limited availability of reliable data pose significant issues, while self-reported measures can introduce biases. Methodologically, many approaches lack generalizability across different populations or outcome measures, requiring domain-specific adjustments and tuning. This field is still evolving, and current studies are largely formative in demonstrating feasibility. Yet real-world applications may introduce new complications. For instance, average accuracy metrics can mask high-stakes risks for individuals in hiring or firing decisions, underscoring the need for novel measurement strategies that consider both risk and qualitative factors. Ensuring rigorous testing of these models—and establishing clear guidelines around algorithmic inference and human judgment—is essential.

Overall, this survey highlights the potential of passive, ubiquitous sensing in the Future of Work, but challenges such as scalability, privacy, and ethical concerns still limit widespread adoption. More research is needed to overcome these barriers and make workplace sensing truly ubiquitous. Our review specifically emphasizes studies published since 2015, which led to the exclusion of older but potentially relevant work. For a broader perspective, we recommend additional surveys ~\cite{Zhang2022,Nappi2020,Vlimki2021,Pramanik2021,Bavaresco2021}. Finally, rapid advances in AI and LLMs offer promising avenues for enhancing workplace interventions via passive sensing. Future work should integrate these technologies to yield interpretable, privacy-preserving insights while upholding fairness and transparency.

\section{Conclusion}
In this survey, we have showcased the latest and most pertinent studies that illuminate current research trends and the promising future of passive sensing technology in the workplace. Much of the research we have discussed is still unfolding, with data-driven investigations paving the way for identifying the most effective sensor streams and models to encapsulate various aspects of the workplace and its workforce. We have also pointed out several unresolved issues and viable pathways for the integration of passive sensing into the Future of Work landscape. As this field is just beginning to bloom, it is intriguing to consider how the integration of advancements in AI and LLMs with passive sensing research could potentially pave the way for innovative developments. While examples within our survey are yet to explicitly showcase this fusion, it is reasonable to anticipate that such advancements may offer promising directions for enhancing workplace productivity and wellbeing in the future.

\bibliographystyle{splncs04}
\bibliography{referencesFull}
%





\end{document}